# The use of controlled vocabularies in requirements engineering activities: a protocol for a systematic literature review.

José L. Barros-Justo[1]        Samuel Sepúlveda[2]
Nelson Martínez-Araujo[1]      Alejandro González-García[1]

[1] School of Informatics (ESEI), University of Vigo, 32004 Ourense, Spain
[2] Departamento de Ciencias de la Computación e Informática (DCI), Centro de Estudios en Ingeniería de Software (CEIS), Universidad de La Frontera, Casilla 54-D, Temuco, Chile.

ABSTRACT

**Context:** The Evidence-Based Software Engineering (EBSE) paradigm and the planning phase of a systematic literature review.

**Objective:** A protocol to do a systematic literature review with detailed information about the processes suggested by several guidelines in the field of evidence-based software engineering.

**Method:** An analisys of recent systematic literature reviews published in world leading journals, plus the use of two renowned guidelines and a textbook to sinthetise a formal plan (the protocol).

**Results:** The validated protocol

**Conclusions:** We found that most of the published systematic reviews lack on reporting the protocol, or it is weak. There is a lack of tool support to develop formal protocols. Although a protocol, like a plan, must have the flexibility to adapt to unforeseen situations, its objective is that the actual activities should resemble as far as possible of those already planned. Therefore, it is a difficult balance to achieve and, researchers must be careful not to introduce alterations that could become threats to the validity of the entire work.

## 1. Introduction.

This document details the planning phase of a Systematic Literature Review (SLR). Our goal is to assess the use of Controlled Vocabularies during the Requirements Engineering phase of software development and, to understand the impact of this usage on different characteristics of the development process (such as productivity and quality). By reviewing the published literature up to March 2017, we will analise what the research community has reported on the application of controlled vocabularies, in academic or industrial contexts, while performing requirements engineering activities, including: the specific vocabularies employed, the engineering activities, the context, the outcomes (positive or negative) and the evidence offered to validate that outcomes. In the following CV stand for Controlled Vocabulary and RE for Requirements Engineering.

As a research method we have chosen a systematic literature review due to its adequacy for the exploration, analysis and synthesis of a research area in a systematic way [1] [2]. Following the advice in [3], [4] and [5], we have decided to develop the protocol for the study as a previous and independent document (see Figure 1). This protocol contain all the details needed to replicate our work by any other researchers and, in doing that, assessing the validity of the work done [6].

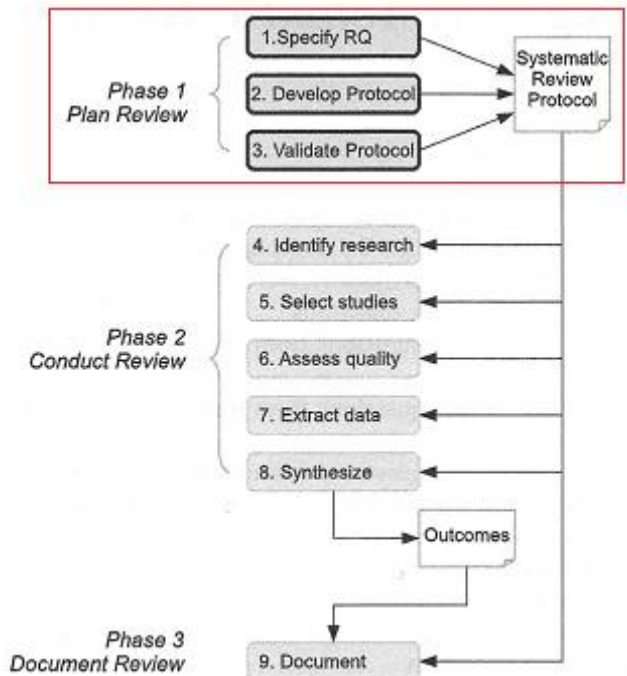

*Figure 1  A systematic review Protocol (Adapted from [5])*

Although any plan or protocol must provide sufficient flexibility to accommodate unexpected situations, we will try, as far as possible, to adjust the execution and reporting of the SLR to the guidelines provided in this document. The authors and two external reviewers agreed this protocol before starting the conducting phase of the SLR.

The following sections present detailed information on how to perform the activities of Phase 2 (Conduct the Review).

## 2. Need for the study.

To set the need of the study we first conducted pilot searches for secondary studies in the area of interest. We ran these searches in the most often used Electronic Data Sources (EDS): ACM Digital Library[1], IEEE Xplore[2], ISI Web of Science[3] and SCOPUS[4]. We used a general search string, composed of some key terms from our area of interest, namely: *controlled vocabularies, requirements engineering activities, reviews, mappings*. None of these searches returned any document, so, as far as we know; there are no secondary studies in the research area prior to this one. The Figure 2 shows the path we followed to justify the need for this study.

---

[1] http://dl.acm.org/
[2] http://ieeexplore.ieee.org/Xplore/home.jsp
[3] https://login.webofknowledge.com
[4] https://www.scopus.com/

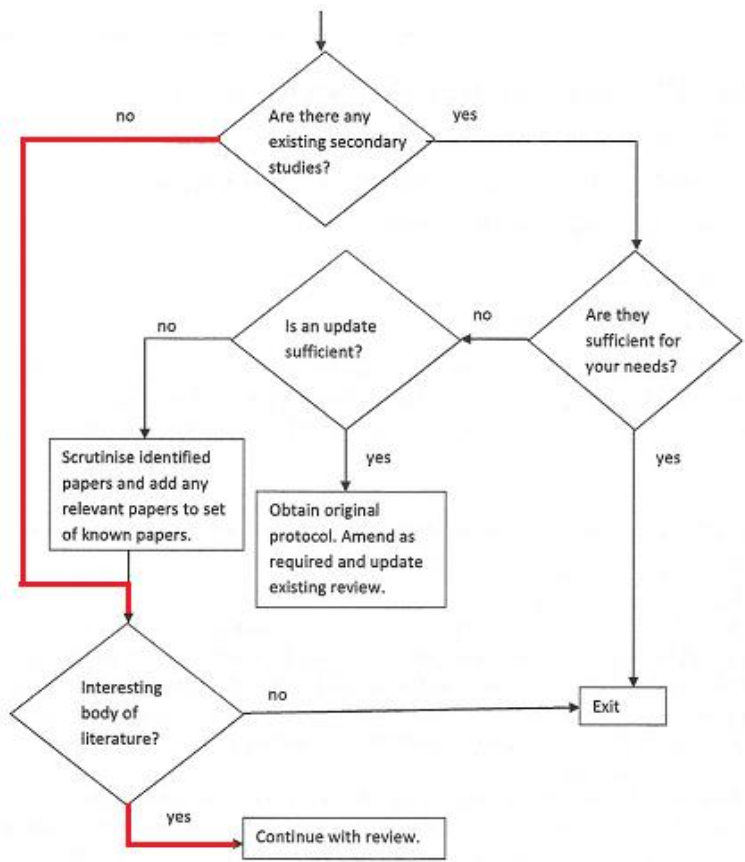

*Figure 2 Need for the study (adapted from [5])*

Our motivation for conducting this systematic review include:
- to gather knowledge about the use of CV in software development,
- to identify the activities of the requirements engineering phase in which CV are used,
- to assess the influence (positive or negative) that the use of CV has on the development process and on the final product,
- to spot potential research gaps and open research lines.

In summary, we are interested in describing and organising the state-of-the-art, by analising and synthesize reported data about the use of CV during the RE activities of software development. We are also interested in assessing potential relationships between CV ←→ RE activities ←→ influence on software development or on the final artifact.

## 3. Goal & RQs.

The main research goal of this systematic literature review is:

- *To assess the use of controlled vocabularies during the requirements engineering phase of software development.*

This goal is broad enough to allow us to slice it in several, more specific, Research Questions (RQs), to address the motivations of the SLR. Table 2 describe the set of RQs considered for this study.

*Table 1 Description of the Research Questions*

| Research Question | Description |
|---|---|
| RQ1: Which features characterize the CV? | What the CV is based on? (e.g., Thesauri, Taxonomies, Glossary, Ontology, Folksonomy…); How can it be accessed? How is it implemented? |
| RQ2: In which RE activities have the CV been used? | The name of the activity, verbatim, as reported in the selected primary works. |
| RQ3: Which aspects of the software development process, or of the final product, were affected by the use of the CV? | The name of the aspect, verbatim, as it appears in the original work (e.g., productivity, quality, development time, ease of maintenance, reduction of bugs…) |
| RQ4: Which features characterize the context where the CV has been used? | - Type of context: Academy or Industry (Domain)<br>- Type of project: toy project, real world application<br>- Development methodology<br>- Duration (time-frame)<br>- Type of requirements: Functional, NonFunctional, Both |
| RQ5: What is the strength of evidence? | - What are the research types?<br>- Which were the research methods?<br>- What is the quality? In terms of rigor and relevance. |

## 4. Search strategies.

Two different and complementary search strategies will be used to ensure, as far as possible, that we will find all available evidence. As suggested by many guidelines on systematic literature reviews [7], [8] and [9] we will use:

- Snowballing (backward and forward).
- Automated search on five different EDS.

### 4.1. Snowballing search

The snowballing search will be conducted according to the guidelines by [10], and will consist of the following steps:

#### 4.1.1. Initial Set selection

The initial set for the snowballing process will consist of some seminal papers recommended by a domain expert. To reduce bias and ensure a broad coverage, these seminal papers will be selected considering the diversity of authors, publication years and venues.

#### 4.1.2. Backward & forward search

For each paper in the initial set, two authors will perform, independently, both backward and forward snowballing, while a third author will combine the results and check for disagreements, as in the paper selection process (see section 5.1).

For backward search, we will review the works listed in the reference section of each paper in the initial set. The reviewers will decide whether each referenced paper should be added for the next snowballing iteration, following the guidelines on section 5. This process will conclude when no papers are added to the selected set.

The forward search (works that cite the one at hand) will be performed using SCOPUS to retrieve the citations, because it asserts[5] to be the largest citation database of peer-reviewed literature. As in backward snowballing, we will select which papers should be considered for the next iteration by applying the exclusion criteria defined in section 5. The forward search will finish when no new papers could be added to the selected set.

For both process (backward and forward) it is crucial to add only relevant papers. A top-down sequential process to select relevant papers for a new iteration, should include:

1. Examine Title, if "excluded" then go to "End (with this paper)".
2. Examine Venue, if it is not peer-reviewed then exclude the paper and go to "End (with this paper)".
3. Examine Abstract, if "excluded" then go to "End (with this paper)".
4. Retrieve the full text and examine other sections such as Introduction, Results or Conclusions, if "excluded" then go to "End (with this paper)".
5. Select paper for the next iteration.
6. End (with this paper).

### 4.2. Automated search

The first step was the selection of the EDS to be used. As many other systematic reviews suggest [11], [12], [13] and [14], we decided to use five different EDS, classified in two different groups:

1. Index engines
   a. SCOPUS              (SCOPUS)
   b. Web of Science      (WoS)
   c. Google Scholar      (GS)
2. Publisher's sites
   a. IEEE Xplore         (IEEE)
   b. ACM Digital Library (ACM)

Experience has shown that, although many of the listed EDS claim to index the same data, they rarely return an equal set of papers given an identical search string [14]. On the other hand, the selected resources cover almost every venue (Conferences, Workshops and Journals) of software engineering field, which means that we can be confident in finding almost all the existing evidence.

Being an independent indexer with a high recall and a very low precision [15], Google Scholar helps reducing publisher bias and improving coverage, at the cost of increased paper reviewing workload. Therefore, we will only consider the first 20 results retrieved. Moreover, the limitations in the interface and the lack of transparency in the algorithm of this EDS [16] [17] advise against using it for large systematic retrieval tasks. For these two reasons, we have used GS as complementary to the four main EDS already mentioned.

---

[5] http://web.archive.org/web/20170301050928/https://www.elsevier.com/solutions/scopus

### 4.2.1. Search string creation (and evolution)
To obtain key terms for the search string we will apply two different strategies:

1. Analysis of our goal and research questions. In this case, our goal was established as: *To assess the use of controlled vocabularies during the requirements engineering phase of software development*. From this goal, we extracted the terms: *controlled vocabularies*, *requirements engineering* and *software development*. These key terms also came from the RQ1, RQ2 and RQ3, respectively. In turn, RQ4 and RQ5 do not contain further key terms to add.

2. Conduct pilot searches: we will run pilot searches on the EDS mentioned before, using the previous key terms, to identify other relevant terms, synonyms and alternative spellings that are frequently used in published literature both, by authors and editors/publishers. We will look for those terms in the Title, Abstract and Author's Keywords of retrieved papers.

The terms extracted will be connected with Boolean operators to construct the final search string and, finally, it will be tailored to the five selected EDS.

### 4.2.2. Validate the search
The search process will be validated by using a quasi-gold standard (QGS), as proposed in [5] and [9]. This QGS will consist of a set of known papers provided by an external expert in the research area. After conducting the automated search, the performance of the search string should be assessed by computing recall, as follows:

$$Recall = \frac{number\ of\ relevant\ studies\ in\ the\ QGS\ found\ by\ the\ automated\ search}{number\ of\ papers\ in\ the\ QGS}$$

The result will be used to establish a threshold: if the Recall is 80% or greater, the search string will be considered valid. Otherwise, another search string should be generated and validated, until a threshold of, at least 80%, is reached. To allow repeatability, the date when the searches were performed should be reported.

## 5. Selection of papers (Inclusion/Exclusion criteria)
The inclusion and exclusion criteria allows selecting the primary papers that focus on the area of interest. The search processes also used the criteria to: a) guide the process of backward and forward snowballing and, b) help to construct the search string for the automated search (EDS).

A paper should be excluded when it fulfill at least one of the following criteria:
1. Objective criteria (minimum bias):
    a. Not written in English,
    b. Not published in a peer-reviewed venue,
    c. Duplicate reports of the same study (consider only the most recent one),
    d. Grey literature (including books, slide presentations, forewords, PhD or master thesis…),
    e. Not a primary study (secondary and tertiary studies, if any, were considered in the Related works section.
2. Subjective criteria (a potential threat to validity):
    a. Not related to the application of controlled vocabularies on software development,
    b. Not related to any requirements engineering phase of software development.

Any paper not excluded by the above criteria will be included in the set of *selected primary papers*.

The application of the exclusion criteria will be done at two different levels:
1. By reviewing the meta-data information (title-abstract-keywords), if this information is not enough to exclude a paper then,
2. Review the full text, particularly the Introduction and Conclusions sections.

Two authors, independently, will carry out the process of paper selection. These authors will produce two sets of pre-selected papers. Another author, will integrate the two previous lists, check for disagreements and, if necessary, eliminate duplicates.

The selected papers will be identified with a code, as follows:
- ASn: paper "n" from the Automated Search
- BSBij or FSBij: paper "j" from iteration "i" from Backward/Forward SnowBalling
- COn: paper "n" from a Conference
- JOn: paper "n" from a Journal

This allows the traceability of each work in the selected set of papers to the search process from which it was retrieved and the venue classification.

### 5.1. Dealing with disagreements

To deal with disagreements we will apply the inclusive criteria A+B+C+D proposed by [18]. Papers classified as "E" will be considered borderline and they will be listed in an Appendix of the SLR. We will exclude a paper only when both reviewers agreed (category "F") or considered the paper as borderline (category "E").

*Table 2  Dealing with disagreements*

|            |           | Reviewer X |           |         |
|------------|-----------|------------|-----------|---------|
|            |           | Include    | Uncertain | Exclude |
| Reviewer Y | Include   | A          | B         | D       |
|            | Uncertain | B          | C         | E       |
|            | Exclude   | D          | E         | F       |

### 5.2. Validation of the selection process

As a validation of the selection process, we will compute the Kappa statistic between pairs of reviewers (that is the reason why we forced the selection process to be conducted by two reviewers independently, as a blind review process) until a value of 0.8 or greater is achieved. The research team will carry out a set of meetings to discuss the papers until the disagreements were resolved.

## 6. Data extraction.

In the data extraction phase the researchers will read the full text of each article accepted for inclusion in the review and, extract the pertinent data using a standardized data extraction/coding form. The data extraction form (DEF) should be designed to extract the data in an objective, explicit and consistent way by all the researchers. Which data to extract was driven by the RQs. Table 3 provides an overview of the data to be extracted, and how the data fields link to the RQs.

*Table 3 Data to be extracted for RQs*

| Research Question | Data to be extracted | Probable location | Classification |
|---|---|---|---|
| RQ1 | The name of all the features mentioned in the study. | Abstract, Introduction, Content. | Not applicable. |
| RQ2 | The name of all the activities or tasks mentioned in the source (belonging to the requirements engineering phase). | Abstract, Introduction, Content. | Standard for Systems and software engineering — Life cycle processes — Requirements engineering [19] consider all the seven activities (from SR2.1 to SR2.7) and:<br>+ Other (not in the standard)<br>+ Not reported |
| RQ3 | The name of the characteristics (of the process or the product), influenced by the use of the CV. | Abstract, Content, Conclusions. | Open list. Report terms used in the source (verbatim). |
| RQ4 | Context characteristics:<br>- Type (domain): Academy or Industry<br>- Type of project: toy project, real application…<br>- Development methodology<br>- Duration (time-frame)<br>- Type of requirements: Functional, NonFunctional, Both | Abstract, Content, Conclusions. | Not applicable. |
| RQ5 | Research type, method and quality (Rigor and Relevance). | Abstract, Content, Conclusions. | - Research type: classification proposed in [20]. See Figure 3.<br>- Research method: classification proposed in [18]. See Figure 4.<br>- Quality assessment: Rigor and Relevance [21]. See section 7. |

Research type classification (T = True, F = False, • = irrelevant or not applicable, R1–R6 refer to rules).

|  | R1 | R2 | R3 | R4 | R5 | R6 |
|---|---|---|---|---|---|---|
| **Conditions** | | | | | | |
| Used in practice | T | • | T | F | F | F |
| Novel solution | • | T | F | • | F | F |
| Empirical evaluation | T | F | F | T | F | F |
| Conceptual framework | • | • | • | • | T | F |
| Opinion about something | F | F | F | F | F | T |
| Authors' experience | • | • | T | • | F | F |
| **Decisions** | | | | | | |
| Evaluation research | ✓ | • | • | • | • | • |
| Solution proposal | • | ✓ | • | • | • | • |
| Validation research | • | • | • | ✓ | • | • |
| Philosophical papers | • | • | • | • | ✓ | • |
| Opinion papers | • | • | • | • | • | ✓ |
| Experience papers | • | • | ✓ | • | • | • |

*Figure 3  Research type [18]*

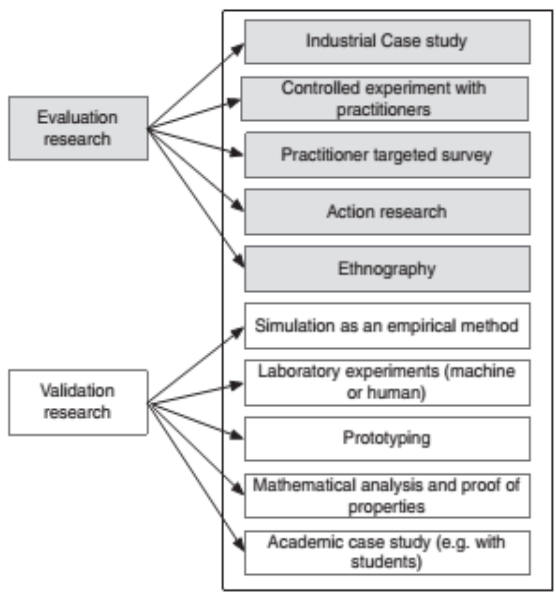

*Figure 4  Research methods and their related research type [18]*

### 6.1. The extraction process

The first author designed a Data Extraction Form (DEF) in spreadsheet format, with columns for every RQ (selected papers will be the rows). The other authors reviewed and agreed the DEF before the extraction process begins. Every single cell should contain:

1. Data extracted (depending on the RQ at hand this data can be a single item or a more complex piece of information)
2. A comment indicating: # of page and the original text (short one) from the source, that justify the data extracted (see Figure 1). If the original text contains other irrelevant information it can be paraphrasing, otherwise, the exact text from the source will be added to the DEF.

To reduce bias, we will divide the set of selected primary papers into two halves:

1. First half: This first half will be assigned to reviewers R1 and R2, without any of them know about the other (blind assignment).
2. Second half: This second half will be assigned to reviewers R3 and R4 (as a blind assignment).

Each pair of reviewers will fill the DEF independently. If conflicts arise then a consensus meeting is held until the disagreements were resolved.

*Figure 5     Data Extraction Form (DEF)*

When a paper presents more than one CV, or more than one RE activity, we will report the data of each item in a new row.

## 7. Quality assessment

Following the guidelines of [21], we will assess the research rigor and industrial relevance of each primary selected study, in order to improve internal and external validity. We will evaluate the rigor as the sum of the context, the study design and the validity descriptions, from a score of 0 to 3 (Table 4). On the other hand, the relevance will be calculated as the sum of the study subjects, context, scale and research method, resulting in a minimum score of zero and a maximum of four (Table 5).

*Table 4 Rubric for Rigor*

| Aspect | Strong description (1) | Medium description (0,5) | Weak description (0) |
|---|---|---|---|
| Context | The context is described in a manner that allows comparison with another. | The context is mentioned, but it is not possible to compare it to another. | There is no description of the context. |
| Study design | The study design is described so that the variables measured, the selection, the control used, etc. are understandable by a reader. | The design is described, but lacks some of the aspects that are necessary for a reader to understand it completely. | There is no description of the study design. |
| Validity description | Different types of validity threats are mentioned, and measures to mitigate them were described. | There are mentions to validity threats, but they are not described in detail. | There is no mention to validity threats. |

*Table 5 Rubric for Relevance*

| Aspect | Contributes (1) | Doesn't contribute (0) |
|---|---|---|
| Subjects | The subjects used in the evaluation are representative of the intended users of controlled vocabularies, i.e. software development practitioners. | The subjects used in the evaluation are not representative of the intended users. |
| Context | The evaluation is performed in a setting that represents the intended usage, i.e. the industry. | The setting where the evaluation is performed is not representative of the intended usage. |
| Scale | The scale of the applications used in the evaluation is realistic, i.e. projects. | The scale of the applications used in the evaluation is not realistic (e.g. toy examples). |
| Research method | The research method used facilitates investigating real situations and is relevant for practitioners (action research, case study, descriptive survey…). | The research method used does not help investigating real situations (e.g. conceptual analysis of CV). |

We agreed with Kitchenham et al. [5] in that "*There is little point in collecting data about primary study quality if you have no plan as to how such data will be used*". We encourage the following two applications of Rigor and Relevance assessment:

- Use rubrics for Rigor and Relevance as part of the inclusion criteria to screen out low quality studies.
- Use the Rigor's score as part of an assessment of the strength of the evidence supporting individual findings.

## 8. Data synthesis and aggregation strategy.

In this section, we define, in advance, the strategy for summarising, integrating, combining and comparing the findings from the primary selected studies. Frequently used approaches to synthesis include narrative and thematic synthesis, where data is tabulated in a way that is consistent with the research questions.

## 9. Threats to validity.

To deal with the potential validity threats we will follow the guidelines in [18]. Therefore, we will consider the following five types of validities:

### 9.1. Theoretical validity.

This is the ability to capture what we intend to capture. This validity is often subdivided into two tasks: identification/selection of studies and data extraction and classification.

Reviewer bias is a significant threat to both paper selection and data extraction, to alleviate its effects two authors will perform the processess independently, and a third will check for disagreements. We will only exclude a paper if both reviewers agree (see the Selection of papers section of this protocol), to minimize the possibility of not taking into account a relevant paper. Additionally, the selection should be validated by establishing a threshold for the Kappa statistic.

## 9.2. Descriptive validity.

It is concerned with the accurate and objective description of data, so it is relevant during the data extraction process. To reduce a potential research bias we will design a Data Extraction Form (DEF) by mutual agreement among all reviewers. The DEF will be implemented as a spreadsheet with columns reflecting the information needs to answer the research questions, and rows for each selected primary work. A general purpose DEF is presented in Figure 5. This DEF should be adapted to fit a specific SLR. Two reviewers, independently, will fill the DEF, while a third reviewer will be responsible of the integration of data and of dealing with possible disagreements.

Every cell in the DEF should contain:

a) the piece of information (data) derived from the original paper, and
b) a comment containing the original text (in the selected primary paper) that supports the reported information, which helps reducing reviewer bias.

## 9.3. Interpretive validity.

It refers to the extent at which every conclusion obtained is justified by an objective analysis of the data collected. Again, researcher bias is a threat to this type of validity, affecting the synthesis processes. To mitigate it, we have established coding rules and, periodic meetings will be held between the authors involved in the data synthesis process to resolve disagreements and ensure that information is interpreted in a consistent way.

## 9.4. Generalizability.

Internal: it occurs within a group, e.g. the same organization. In our SLR, internal generalizability is determined by the CV usages that were reported in previous studies. Hence, a large number of primary studies reporting the results of using CV, in different ways or contexts, will benefit the internal generalizability while a low number will be a threat.

External: it occurs within groups, e.g. between organizations. We cannot estimate the external generalizability (different CV usages to the ones that were subjects of previous studies) during the planning phase, so that could be investigated while conducting the review.

## 9.5. Reliability.

It refers to the repeatability of the research process. This repeatability requires detailed reporting of the research process; that is the main purpose of this document (the protocol itself). All the search, paper selection, data extraction and synthesis processes will be conducted according to the guidelines specified in the other sections of this protocol, so the conducting phase will have a great degree of repeatability, providing that no need to do significant changes to the protocol arise. For example, reporting the date when the searches were performed, or indicating how a piece of information in the DEF link to a piece of text in the original source, both, contribute to increase the reliability of our study.